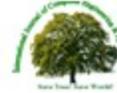

# HAPTIC SCIENCE AND TECHNOLOGY

S. Sri Gurudatta Yadav, Research Scholar, DRKGI, Hyderabad, India.

*Dr.R.V.Krishnaiah*, PG-Coordinator, DRKGI, Hyderabad, India.

**Abstract:**

Haptic technology, or haptics, is a tactile feedback technology which takes advantage of a user's sense of touch by applying forces, vibrations, and/or motions upon the user. This mechanical stimulation may be used to assist in the creation of virtual objects (objects existing only in a computer simulation), for control of such virtual objects, and for the enhancement of the remote control of machines and devices. It has been described as "for the sense of touch what computer graphics does for vision". Although haptic devices are capable of measuring bulk or reactive forces that are applied by the user, it should not be confused with touch or tactile sensors that measure the pressure or force exerted by the user to the interface.

**Keywords:** Tactile, Virtual Objects Creation and Control, Sense of touch, Haptic Rendering, Haptic Perception.

### I. INTRODUCTION:

"Haptic", is the term derived from the Greek word, "haptesthai", which means 'sense of touch'. Haptic is defined as the "science of applying tactile sensation to human interaction with computers". Haptic permits users to sense ("feel") and manipulate three-dimensional virtual objects with respect to such features as shape, weight, surface textures, and temperature.

By using Haptic devices, the user can not only feed information to the computer but can receive information from the computer in the form of a felt sensation on some part of the body. This is referred to as a Haptic interface [1].

In our paper we explain the basic concepts of 'Haptic Technology and its Application in Surgical Simulation and Medical Training'. PHANTOM is small robot arm with three revolute joints each connected to a computer-controlled electric DC motor.

Cyber Grasp is used in conjunction with a position tracker to measure the position and orientation of the fore arm in three-dimensional space. Phantom and Cyber Grasp are Hapic devices [4].

### II. LITERATURE REVIEW:

Force feedback is the area of haptics that deals with devices that interact with the muscles and
Tendons that give the human a sensation of a force being applied with hardware and software that stimulates humans' sense of touch and feel through tactile vibrations or force feedback.
These devices mainly consist of robotic manipulators that push back against a user with the forces that correspond to the environment that the virtual effector's is in.

Tactile feedback makes use of devices that interact with the nerve endings in the skin to indicate heat, pressure, and texture.

These devices typically have been used to indicate whether or not the user is in contact with a virtual object. Other tactile feedback devices have been used to stimulate the texture of a virtual object.

**Human Senses:**

Typically, it is believed that vision and audition convey the most information about an environment while the other senses are more subtle. Because of this, their characteristics have been widely investigated over the last few decades by scientists and engineers, which has led to development of reliable multimedia systems and environments [6]. Haptic input device is shown in the following fig1.

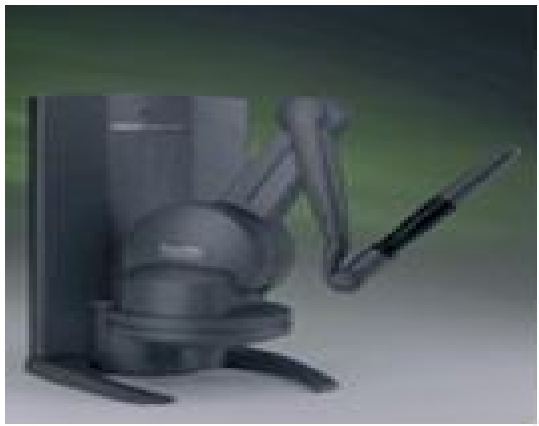

Fig1: Haptic Device

**Vision:**

The visual sense is based on the level of absorption of light energy by the eye and the conversion of this energy into neural messages. The acceptable wavelength range for human eyes is between 0.3 and 0.7m. The temporal resolution sensitivity of the human visual system is biologically limited and not sufficient to detect the presentation of sequential video frames past a certain speed. Similarly, our spatial resolution is limited and does not allow us to resolve individual pixels.

**Audition:**

The human auditory system transmits sound waves through the outer, middle, and inner ears. This sound wave is transformed into neural energy in the inner ear. It is then transmitted to the auditory cortex for processing. The audible frequency of humans ranges from 16 to 20,000Hz and is most efficient between 1,000 and 4,000Hz.

**Touch:**

Indeed, the sense of touch is distributed over the entire body, unlike the other conventional four senses, which are centralized around specific parts of the body. The sense of touch is mainly associated with active tactile senses such as our hands. To appreciate the sense of touch more fully, consider the following facts: according to Heller and Schiff, touch is twenty times faster than vision, so humans are able to differentiate between two stimuli just 5ms apart; Bolanowski et al. found that touch is highly sensitive to vibration up to 1KHz, with the peak sensitivity around 250 Hz; and skin receptors on the human palm can sense displacements as low as 0.2_m in length.

### III. HAPTIC TECHNOLOGIES

Tactile cues include textures, vibrations, and bumps, while kinesthetic cues include weight, impact, etc. In the following section, we present some crucial concepts and terminology related to haptics.

**Haptic:** the science of applying tactile, kinesthetic, or both sensations to human computer interactions. It refers to the ability of sensing and/or manipulating objects in a natural or synthetic environment using a haptic interface.

**Cutaneous:** relating to or involving the skin. It includes sensations of pressure, temperature, and pain.

**Tactile:** pertaining to the cutaneous sense, but more specifically the sensation of pressure rather than temperature or pain.

Kinesthetic: relating to the feeling of motion. It is related to sensations originating in muscles, tendons, and joints.

**Force Feedback:** relating to the mechanical production of information that can be sensed by the human kinesthetic system.

**Haptic communication:** the means by which humans and machines communicate via touch. It mostly concerns networking issues.

**Haptic device:** is a manipulator with sensors, actuators, or both. A variety of haptic devices have been developed for their own purposes. The most popular are tactile-based, pen-based, and 3 degree-of-freedom (DOF) force feedback devices.

**Haptic interface:** consists of a haptic device and software-based computer control mechanisms. It enables human–machine communication through the sense of touch. By using a haptic interface, someone can not only feed the information to the computer but can also receive information or feedback from the computer in the form of a physical sensation on some parts of the body.

**Haptic perception:** the process of perceiving the characteristics of objects through touch.

**Haptic rendering:** the process of calculating the sense of touch, especially force. It involves sampling the position sensors at the haptic device to obtain the user's position within the virtual environment. Haptic rendering is, therefore, a system that consists of three parts, a collision detection algorithm, a collision response algorithm, and a control algorithm [3].

**Sensors and Actuators:** a sensor is responsible for sensing the haptic information exerted by the user on a certain object and sending these force readings to the haptic rendering module. The actuator will read the haptic data sent by the haptic rendering module and transform this information into a form perceivable by human beings [6].

**Tele-haptics:** the science of transmitting haptic sensations from a remote explored object/environment, using a network such as the Internet, to a human operator. In other words, it is an extension of human touching sensation/capability beyond physical distance limits.

**Tele-presence:** the situation of sensing sufficient information about the remote task environment and communicating this to the human operator in a way that is sufficient for the operator to feel physically present at the remote site. The user's voice, movements, actions, etc. may be sensed, transmitted, and duplicated in the remote location. Information may be traveling in both directions between the user and the remote location [2].

**Virtual Reality (VR):** can be described as the computer simulation of a real or virtual (imaginary) world where users can interact with it in real time and change its state to increase realism. Such interactions are sometimes carried out with the help of haptic interfaces, allowing participants to exchange tactile and kinesthetic information with the virtual environment.

**Virtual environment (VE):** is an immersive virtual reality that is simulated by a computer and primarily involves audiovisual experiences. Despite the fact that the terminology is evolving, a virtual environment is mainly concerned with defining interactive and virtual image displays.

**Collaborative virtual environments (CVE):** is one of the most challenging fields in VR because the simulation is distributed among geographically dispersed computers. Potential CVE applications vary widely from medical applications to gaming.

**Collaborative haptic audio visual environment (C-HAVE):** in addition to traditional media, such as image, audio, and video, haptics – as a new media – plays a prominent role in making virtual or real-world objects physically palpable in a CVE. A C-HAVE allows multiple users, each with his/her own haptic interface, to collaboratively and/or remotely manipulate shared objects in a virtual or real environment and is shown in fig2.

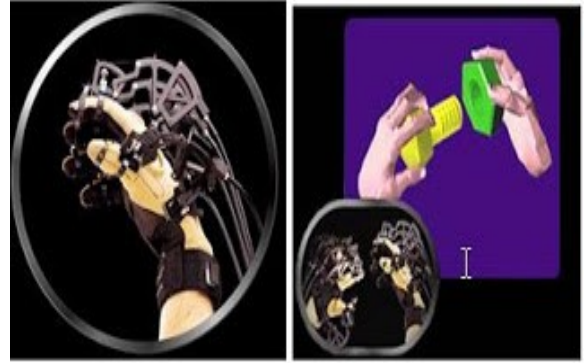

Fig2: Haptic Visual Environment

It is a process of applying forces to the user through a force-feedback device. Using haptic rendering, we can enable a user to touch, feel and manipulate virtual objects. Enhance a user's experience in virtual environment. Haptic rendering is process of displaying synthetically generated 2D/3D haptic stimuli to the user. The haptic interface acts as a two-port system terminated on one side by the human operator and on the other side by the virtual environment[7].

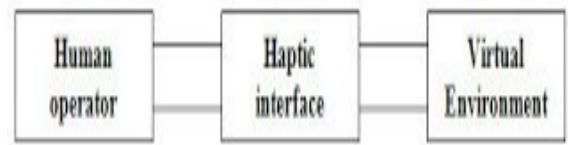

Fig3: Haptic structure

**Contact Detection:**

A fundamental problem in haptics is to detect contact between the virtual objects and the haptic device (a PHANTOM, a glove, etc.). Once this contact is reliably detected, a force corresponding to the interaction physics is generated and rendered using the probe. This process usually runs in a tight servo loop within a haptic rendering system.

**Surgical Simulation and Medical Training:**

Haptic is usually classified as:-

**Human haptics:** human touch perception and manipulation.

**Machine haptics:** concerned with robot arms and hands.

**Computer haptics:** concerned with computer mediated.

A primary application area for haptics has been in surgical simulation and medical training. Haptic rendering algorithms detect collisions between surgical instruments and virtual organs and render organ-force responses to users through haptic interface devices. For the purpose of haptic rendering, we've conceptually divided minimally invasive surgical tools into two generic groups based on their functions.

1. Long, thin, straight probes for palpating or puncturing the tissue and for injection (puncture and injection needles and palpation probes)

2. Articulated tools for pulling, clamping, gripping, and cutting soft tissues (such as biopsy and punch forceps, hook scissors, and grasping forceps).

Haptic Science operational structure is shown in following fig4.

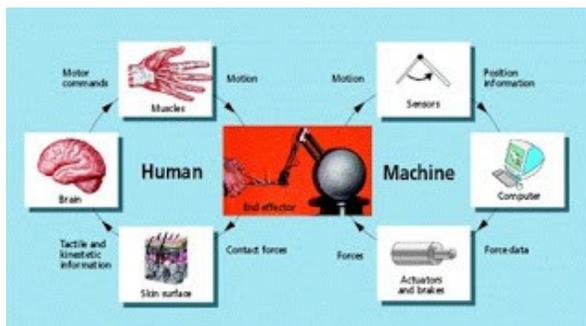

Fig4: Haptic Science operational structure

Grouping of surgical instruments is used for simulating tool tissue interactions. Group A includes long, thin, straight probes. Group B includes tools for pulling, clamping, and cutting soft tissue [7] which is shown in fig5.

1. Users feel torques if a proper haptic device is used. For example, the user can feel the coupling moments generated by the contact forces at the instrument tip and forces at the trocar pivot point.

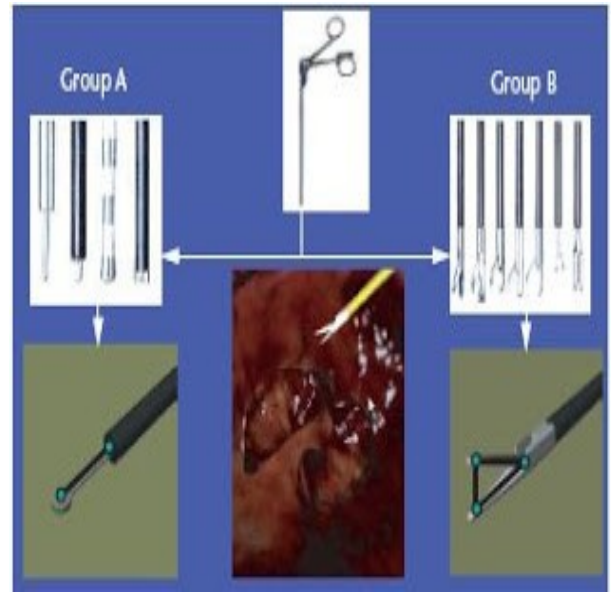

Fig5: Haptic device Interaction

2. Users can detect side collisions between the simulated tool and 3D models of organs.
3. Users can feel multiple layers of tissue if the ray representing the simulated surgical probe is virtually extended to detect collisions with an organ's internal layers.
4. Users can touch and feel multiple objects simultaneously. Because laparoscopic instruments are typically long slender structures and interact with multiple objects (organs, blood vessels, surrounding tissue, and so on) during a MIS (Minimally Invasive Surgery), ray-based rendering provides a more natural way than a purely point-based rendering of tool-tissue interactions.

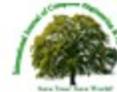

To simulate haptic interactions between surgical material held by a laparoscopic tool (for example, a catheter, needle, or suture) and a deformable body (such as an organ or vessel), a combination of point- and ray-based haptic rendering methods are used which are shown in fig6 [8].

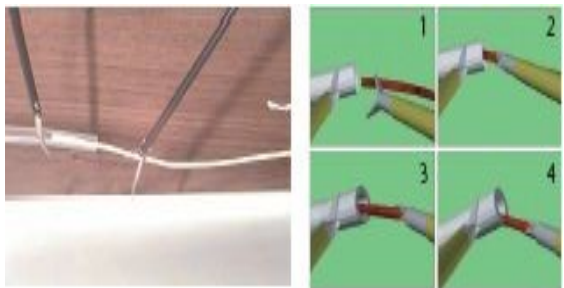

Fig6: Haptic rendering structure

In case of a catheter insertion task shown above, the surgical tools using line segments and the catheter using a set of points uniformly distributed along the catheter's center line and connected with springs and dampers. Using our point based haptic rendering method; the collisions between the flexible catheter and the inner surface of a flexible vessel are detected to compute interaction forces.

**Deformable Objects:**
One of the most important components of computer based surgical simulation and training systems is the development of realistic organ-force models. A good organ-force model must reflect stable forces to a user, display smooth deformations, handle various boundary conditions and constraints, and show physics-based realistic behavior in real time [5].

Tool–tissue interactions generate dynamical effects and cause nonlinear contact interactions of one organ with the others, which are quite difficult to simulate in real time. Furthermore, simulating surgical operations such as cutting and coagulation requires frequently updating the organ geometric database and can cause force singularities in the physics-based model at the boundaries. There are currently two main approaches for developing force-reflecting organ models:

1. Particle-based methods.
2. Finite-element methods (FEM).

In particle-based models, an organ's nodes are connected to each other with springs and dampers. Each node (or particle) is represented by its own position, velocity, and acceleration and moves under the influence of forces applied by the surgical instrument.

**Capture, Storage, and Retrieval of Haptic Data:**
The newest area in haptic is the search for optimal methods for the description, storage, and retrieval of moving-sensor data of the type generated by haptic devices. This techniques captures the hand or finger movement of an expert performing a skilled movement and "play it back," so that a novice can retrace the expert's path, with realistic touch sensation; The INSITE system is capable of providing instantaneous comparison of two users with respect to duration, speed, acceleration, and thumb and finger forces.

Techniques for recording and playing back raw haptic data have been developed for the PHANTOM and Cyber Grasp. Captured data include movement in three dimensions, orientation, and force (contact between the probe and objects in the virtual environment) [6].

**Haptic Datacompression:**

Data about the user's interaction with objects in the virtual environment must be continually refreshed if they are manipulated or deformed by user input. If data are too bulky relative to available bandwidth and computational resources, there will be improper registration between what the user sees on screen and what he "feels."

On analyzing data obtained experimentally from the PHANTOM and the Cyber Grasp, exploring compression techniques, starting with simple approaches (similar to those used in speech coding) and continuing with methods that are more specific to the haptic data. One of two lossy methods to compress the data may be employed: One approach is to use a lower sampling rate; the other is to note small changes during movement.

## IV. ROADMAP TO MULTIMEDIA HAPTICS:

In a virtual environment, a real scenario is simulated by a computer generated application where some of the user's senses are ingeniously represented in order for them to interact and perceive stimuli that are very similar to the real environment. Traditionally, human–computer interfaces have delivered types of stimuli that are based on two of our senses, namely vision and sound. However, with the addition of the sense of touch through tactile and force feedback, the computer-based applications become richer in media content through better mimicry of real-life situations and tasks or remote real environments. The sensing of forces is tightly coupled with both the visual system and one's spatial sense; the eyes and hands work collectively to explore and manipulate objects. Moreover, researchers have demonstrated that haptic modality reduces the perceived musculoskeletal loading that is measured through pain and discomfort in completing a task. Therefore, there is a trend in the design of interfaces toward multimodal human–computer interaction that incorporates the sense of touch.

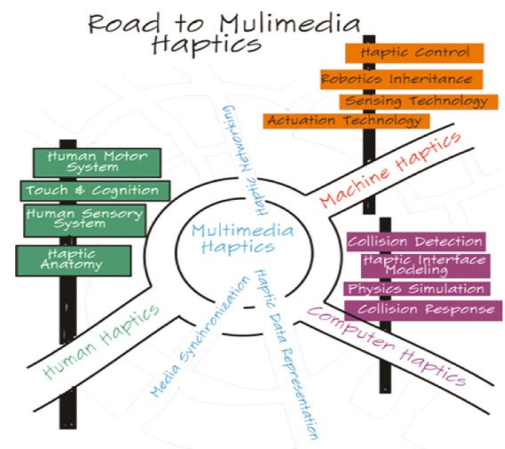

Fig7: Roadmap to multimedia haptics

However, the keyword here is "perception", so if the cross-modal information is not well synchronized and consistent, the added sensory information might corrupt the intended stimulus. For instance, researchers have found that when conflict between sensory cues (for instance, between the hands and eyes) arise, the brain effectively splits the difference to produce a single mental image, and the overall perception experienced by the subject will be a compromise between the two senses. Therefore, visual cues must be synchronized with haptic interactions to increase the quality of perception [9].

**Applications of Haptic Technology:**

Haptic Technology applications are Surgical simulation & Medical training, Physical rehabilitation, Training and education, Museum display, Painting, sculpting and CAD, Scientific Visualization, Military application and Entertainment.

### V. RESULTS:

We conclude that Haptic Technology is the only solution which provides high range of interaction that cannot be provided by virtual reality. The touch access technology is important till now. But, haptic technology has totally changed this trend. This technology make the future world as a sensible one. Haptic Technology enables users to simulate touch and utilize a new input as well as output technology Large potential for applications in critical fields as well as for leisurely pleasures. Haptic devices must be miniaturized so that they are lighter, simpler and easier to use.

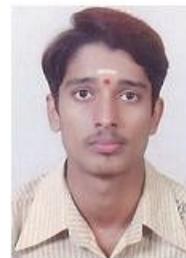
**S. Sri Gurudatta Yadav**, M.Tech in CSE from JNTU Hyderabad.

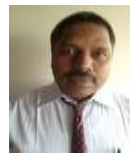
**Dr. R. V. Krishnaiah** has received Ph.D from JNTU Ananthapur, M.Tech in CSE from JNTU Hyderabad, M.Tech in EIE from NIT (former REC) Warangal and B.Tech in ECE from Bapatla Engineering College. He has published more than 50 papers in various journals. His interest includes Data Mining, Software Engineering, Image Processing.